\documentclass[reprint,showpacs,preprintnumbers,amsmath,amssymb, aip]{revtex4-1}
\usepackage{graphicx,wasysym,color}
\usepackage{dcolumn}
\usepackage{bm}
\usepackage{tabulary}

\newcommand{\Tm}{T_{\mathrm{m}}}

\newcommand{\Ts}{T_{\mathrm{s}}}

\newcommand{\dd}{\mathrm{d}}

\newcommand{\kB}{k_\mathrm{B}}

\begin{document}

\preprint{1}

\title{Premelting, fluctuations and coarse-graining of water-ice interfaces}

\author{David T. Limmer}
 \email{dlimmer@princeton.edu}
\affiliation{%
Princeton Center for Theoretical Science, Princeton University, Princeton, NJ, USA 08540
}
\author{David Chandler}
\affiliation{%
Department of Chemistry, University of California, Berkeley, CA, USA 94609
}%

\date{\today}
\begin{abstract}
Using statistical field theory supplemented with molecular dynamics simulations, we consider premelting on the surface of ice as a generic consequence of broken hydrogen bonds at the boundary between the condensed and gaseous phases. A procedure for coarse-graining molecular configurations onto a continuous scalar order parameter field is discussed, which provides a convenient representation of the interface between locally crystal-like and locally liquid-like regions. A number of interfacial properties are straightforwardly evaluated using this procedure such as the average premelting thickness and surface tension.  The temperature and system size dependence of the premelting layer thickness calculated in this way confirms the characteristic logarithmic growth expected for the scalar field theory that the system is mapped onto through coarse-graining, though remains finite due to long-ranged interactions. Finally, from explicit simulations the existence of a premelting layer is shown to be insensitive to bulk lattice geometry, exposed crystal face and curvature.
\end{abstract}

\maketitle

A premelting layer refers to thermodynamically stable disordered structure at the interface of an otherwise ordered crystalline solid.  It appears when temperature is near but below the bulk melting temperature.\cite{nelson1989statistical} First proposed by Michael Faraday in 1842 to explain the low friction of the surface of ice,\cite{faraday1859athenaeum} definitive experimental evidence for surface melting was not observed on ice until 1987.\cite{kouchi1987x} Since then advances in surface selective experimental techniques have provided powerful tools for direct atomic-resolution observations of this surface phase transition.\cite{li2007surface} These experimental studies have been complemented by a number of detailed atomistic simulations that also find a premelting layer on the surface of models of ice.\cite{bishop2009thin,watkins2011large,neshyba2009molecular,conde2008thickness,shepherd2012quasi} Despite these previous studies, at present no microscopic description exists that both establishes the precise nature of the transition and connects this level of detail with a more coarse-grained picture necessary to explain experimental observations. Using a simple field theory described previously,\cite{limmer2012phase} complimented with molecular dynamic simulations, we address this deficiency. 
\begin{figure}[b]
\begin{center}
\includegraphics{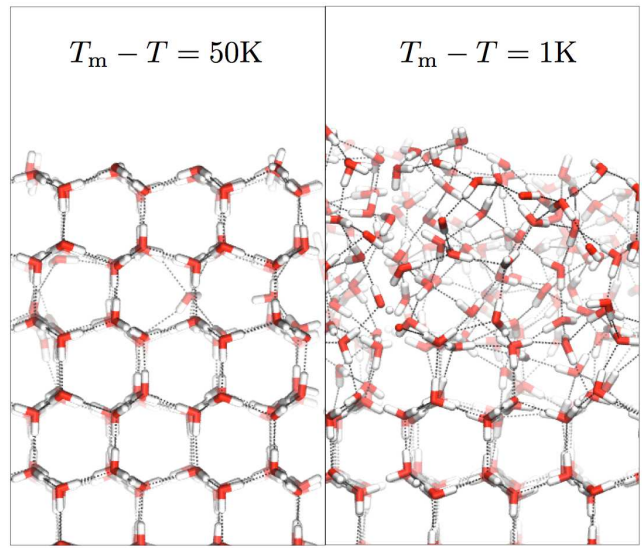}
\caption{Characteristic snapshots of the basal surface of Ice Ih 
taken from our molecular dynamics simulations of the TIP4P/2005 model. }
\label{Fi:1}
\end{center} 
\end{figure}

Most liquids at ambient conditions are close to their triple point and as a consequence, at ambient conditions the chemical potential differences between the liquid, solid and vapor phase are all very small. Therefore, near the melting temperature, $\Tm$, a thermodynamic criterion for the existence of the premelting layer is given by balancing the different surface terms, 
\begin{equation}
\Delta \gamma = \gamma_{s,v} - \gamma_{\ell,s} - \gamma_{\ell,v}   \, , 
\end{equation}
where $\gamma_{s,v}, \gamma_{\ell,s},$ and $\gamma_{\ell,v}$ are the solid-vapor, liquid-solid, and liquid-vapor surface tensions, respectively. When $\Delta \gamma>0$, there is a thermodynamic driving force for premelting. In the case of water and ice, this driving force can be easily rationalized from a microscopic perspective. Figure \ref{Fi:1} shows representative configurations of the surface of ice at conditions far away from, and close to, the melting temperature. To capture these configurations, we have carried out molecular simulations of the TIP4P/2005 model of water.~\cite{abascal2005general} At very cold conditions, i.e, temperature $T$ much below the melting temperature $\Tm$, the surface is ordered and molecules at the surface are forced to break one hydrogen bond on average.  At higher temperatures, $T \rightarrow \Tm$, hydrogen bonding is still disrupted but this energetic loss is balanced by an entropic gain in restoring translational invariance parallel to the surface. The concomitant enhancement of fluctuations at the interface increases in scale as the melting temperature is approached. Exactly at $\Tm$, the degeneracy in the free energy dictates that for an infinite system the thickness of the interface may diverge as it is equally likely that the bulk is liquid or crystal. 

How the thickness of the premelting layer changes with temperature reflects the interplay between bulk free energies that favor order and the boundary condition that excludes it. As reviewed in Ref.~\onlinecite{li2007surface}, experimental estimates of the thickness of the premelting layer at a prescribed temperature, typically $T \approx \Tm-1$\,K, vary by over two orders of magnitude depending on the technique and interpretation. Partially, this variability occurs because different techniques probe different physical properties with differing correlations to structural disorder. X-ray and proton scattering are typically more sensitive to long ranged order, and other surface selective techniques, such as sum frequency generation spectroscopy\cite{wei2001surface} and atomic force microscopy,\cite{bluhm2000friction} have also been used with success. In all cases, however, the strong temperature dependence of the thickness adds to uncertainties due to contamination and surface preparation. Further, in all cases it is not straightforward to connect the observations to molecular level details.

Theory and simulation offer a way to remove this ambiguity. Specifically, using sufficiently general statistical mechanical arguments, which relate experimental observables to emergent behavior with assumptions tested with explicit molecular simulations, the phenomenology of the premelting layer can be understood and quantified. In what follows, we will show that a simple extension of the field theory we have used previously\cite{limmer2012phase} accurately predicts the existence and scaling of the premelting layer as a function of both temperature and system size. The accuracy of our field theory is confirmed with efficient molecular dynamics simulations of the mW model.~\cite{Molinero:2009p4008}

\begin{figure*}[t]
\begin{center}
 \includegraphics{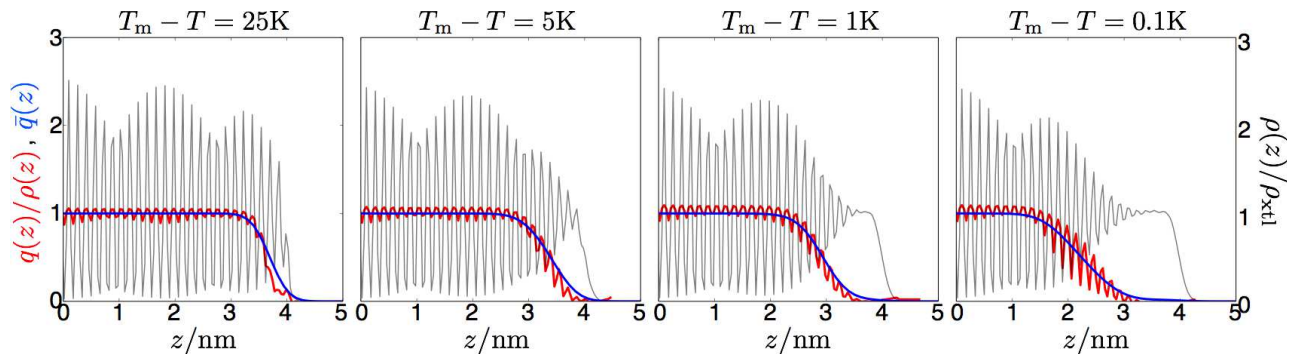}
\caption{Density and order parameter profiles for temperatures away from $\Tm$ computed for the 100 surface of ice Ic of the mW model. Grey lines identify the mean density. Red and blue lines identify the mean order density, $q(z)$, and its coarse-grained analogue, $\bar{q}(z)$, respectively. See text for details.}
\label{equPD:profiles}
\end{center} 
\end{figure*}

\section{Local order parameter profiles}
We begin to examine the premelting layer using molecular dynamics simulations of a minimal model of water, the mW model.\cite{Molinero:2009p4008} This model has been shown previously to accurately model water in the condensed phase, correctly recovering equilibrium liquid\cite{Molinero:2009p4008,Limmer:2011p134503,holten2013nature} and crystal properties\cite{moore2011structural,jacobson2009thermodynamic,moore2011cubic,li2011homogeneous} as well as nonequilibrium and dynamical properties.\cite{limmer2013corresponding,limmer2014theory}

Simulations of ice slabs are set up using the following protocol. First, a perfectly crystalline ice lattice is created with an equilibrium geometry consistent with zero pressure and low temperatures. This lattice is placed in a simulation box and periodically replicated in the $x$- and $y$-direction. The boundary conditions in the $z$-direction are inhomogeneous. For the surface pointing in the positive $z$-direction, we have an open boundary condition, where we expose the surface to its vapor. For the surface pointing in the negative $z$-direction, we attach harmonic restoring forces, with spring constants $k=20~\kB T/$\AA$^2$, to the crystalline lattice positions of the first two layers of water molecules to preserve a crystalline boundary condition. This constraint is not unique but is chosen for convenience in that it is strong enough to maintain crystalline order but not so stiff as to require significantly smaller integration timesteps relative to those typically used in the simulation of the mW model.\cite{Molinero:2009p4008} The system size is characterized by two lengths, the width of the crystal slab, $W$, and its length, $L$. The width is approximately 3 nm and we vary the length between 3 and 13 nm.  The temperature is controlled with a Nose-Hoover thermostat, with a time constant of 1 ps. Depending on the proximity to the melting temperature, simulations were run between 10 and 1000 ns in order to obtain converged estimates of the surface properties. The long timescales are required because the size of fluctuations in the interface become as large as the system as $\Tm$ is approached. Use of the mW model allows us to easily access these timescales. Molecular dynamics calculations are performed with LAMMPS.\cite{plimpton1995fast}

As expected from experiment, the average configuration of the ice slab depends sensitively on temperature. Figure~\ref{equPD:profiles} shows the mean density projected along the direction perpendicular to the plane of the surface. The mean density is calculated by binning the particle positions along the $z$ axis. The density profile is normalized by its mean value in the center of the slab, $\rho_\mathrm{xtl}$, as averaged over many unit cells. Near the center of the slab, the density exhibits the expected oscillations characteristic of crystal structure. At low temperatures, shown in panels (a,b), these oscillations persist to the end of the slab. At high temperatures, $T \rightarrow \Tm$, shown in panels (c,d), these oscillations decay upon approaching the exposed crystal surface. The uniformity of the density near the crystal surface is an indication that the surface is disordered.

For water, locally liquid-like states can be distinguished from more ordered crystal-like states in terms of an order parameter, $q$, that is real and a scalar.  There are many such measures suitable for this purpose that have been used previously.\cite{moore2010freezing,reinhardt2012local,ten1995numerical}  As with our previous work,\cite{limmer2012phase} we chose the local order parameter from Steinhardt, Nelson and Ronchetti,\cite{steinhardt1983bond}  
\begin{equation}
q(\mathbf{r}) + q_\mathrm{liq} = \sum_{i=1}^{N} q^{(i)} \, \delta(\mathbf{r} - \mathbf{r}_i)\,,
\label{eq:orderparameter}
\end{equation}
where $\mathbf{r}_i$ is the position of the $i$th oxygen among $N$ water molecules, and 
\begin{equation}
q^{(i)} = \frac{1}{4}\left( \sum_{m=-6}^{6}\,\, \Big| \sum_{\,j\in \mathrm{nn}(i)} q_{6m}^{(j)} \,\,\Big|^2   \right)^{1/2}\,,
\label{eq:q6i}
\end{equation}
with
\begin{equation}
q_{6m}^{(i)} = \frac{1}{4} \sum_{\,\,j \in \mathrm{nn}(i)} Y_{6m} (\phi_{ij},\theta_{ij})\,.
\label{eq:q6mi}
\end{equation}
Here, the sum over $j\in \mathrm{nn}(i)$ includes only the 4 nearest neighbor oxygens of the $i$th oxygen, and  $Y_{l m} (\phi_{ij} , \theta_{ij})$ is the $l$-$m$ spherical harmonic function associated with the angular coordinates of the vector $\mathbf{r}_i - \mathbf{r}_j$.
  The quantity  $q_\mathrm{liq}$ is the average value of the right-hand-side of Eq.~\ref{eq:orderparameter} for the bulk liquid.  At ambient conditions, $\langle q^{(i)}\rangle_\mathrm{liq} \approx 0.2$.  In contrast, for crystal ice configurations, $\langle q^{(i)}\rangle_\mathrm{xtl} \approx 0.45 $, with root-mean-square fluctuations $\langle (\delta q^{(i)})^2\rangle^{1/2}_\mathrm{xtl} \approx 0.02 $.  As such, and as past experience has shown,\cite{Limmer:2011p134503} using the $l = 6$ spherical harmonics with 4 nearest neighbors discriminates local ice-like structure from typical liquid structure. 

The disorder at the surface can be confirmed by investigating the local order parameter density. The order parameter density is calculated by binning the parameter defined in Eq.~\ref{eq:orderparameter} normalized by its average value in the crystal, 
\begin{equation}\label{eq:qofz}
q(z) = \langle q(\mathbf{r}) \, \delta (z-\hat{z} \cdot \mathbf{r} ) \rangle / q_\mathrm{xtl} \,,  
\end{equation}
where the angle brackets denote equilibrium average, $\hat{z}$ is the unit vector in the $z$-direction, and $q_\mathrm{xtl}$ is the average of $q(\mathbf{r})$ for the crystal. These profiles also oscillate with the periodicity of the lattice, however their amplitudes decay to 0 for $z$ values smaller than the density distributions. For temperatures close to $\Tm$ the difference in the position of decay of the two profiles is greater than 1 nm, while at low temperatures the difference in the position is only on the order of a molecular diameter. 
\begin{figure}
\begin{center}
\includegraphics{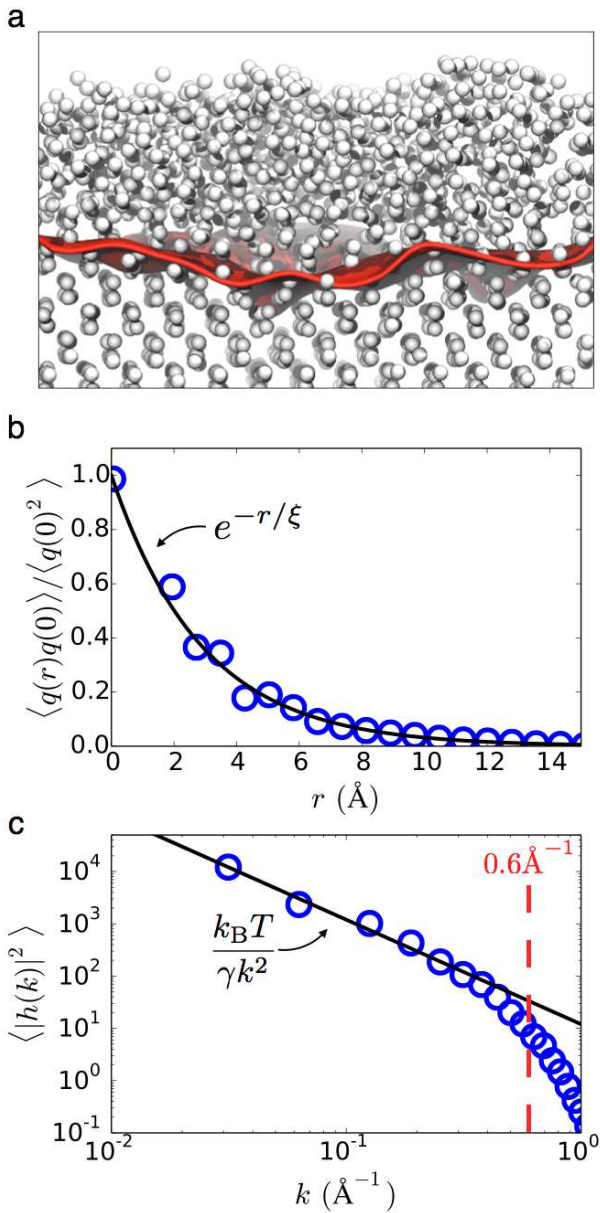}
\caption{Structural correlations in the liquid and interface and the instantaneous local order interface. a) A typical configuration of the premelting layer taken from molecular dynamics simulations of the mW model at $T=272$~K, $p=1$~bar. The membrane, shown in red, locates the isodensity 1/2 surface for the local decay of the coarse-grained order parameter field, $\bar{q}(\mathbf{r})$. b) Order parameter correlation function computed from molecular dynamics simulations of the mW model at $T=273$~K, $p=1$~bar compared with an exponential of characteristic length, $\xi=2.5\mathrm{\AA}$. c) Height-height correlation function for the interface defined by the coarse-grained order parameter field, for the mW model at $T=255$~K, $p=1$~bar. See text for details. }
\label{fig:premeltprop}
\end{center} 
\end{figure}

In principle, this gap between the decay of the density and order density distributions provides a way for determining the length of the premelting layer, $\ell$. However, any simple definition of the decay of the order parameter profile yields a measure of the premelting thickness that is not single valued because of the oscillations in the density and order density. In the next section we show how coarse-graining out these short-wavelength oscillations yields a smooth profile with a well-defined premelting length and produces an effective mapping from this atomistic representation to a scalar field, which can be analyzed simply within a mean field theory. 

\section{Coarse-graining local order}
To remove the short-wavelength components of the profiles in Fig.~\ref{equPD:profiles}, and make contact with the long-wavelength description supplied by the field theory used later, we must integrate out lengthscales small compared to typical intermolecular distances.  A way to do this in a computer simulation is evident from Eq.~\ref{eq:orderparameter}. Rather than using a delta function to construct a continuous field from the local order parameter, we can convolute it with a bounded function of finite width to coarse-grain the local field over molecular distances. We use
\begin{equation}
\label{eq:CG}
\bar{q}(\mathbf{r}) = \int_{\mathbf{r}'} q(\mathbf{r}') \phi(\mathbf{r}'-\mathbf{r};\xi) \,,
\end{equation}
where
\begin{equation}
\phi(\mathbf{r};\xi) = \left (\frac{1}{2\pi \xi^2}\right )^{3/2} e^{-\mathbf{r}^2/2\xi^2}
\end{equation}
introduces the coarse-graining lengthscale $\xi$.
We want $\xi$ to be on the order of the size of the molecule in order to remove the short lengthscale oscillations. In order to determine a reasonable value for $\xi$, we adopt the estimate
\begin{equation}
\xi = \int_0^\infty \,\dd r \, \langle q(r) q(0)\rangle / \langle q^2\rangle \, ,
\end{equation}
where the ensemble average is carried out in the liquid.  The correlation function is plotted in Fig.~\ref{fig:premeltprop} for a simulation of 8000 mW molecules at $T=273$~K and $p=1$~bar. This integral yields $\xi=2.5$\,\AA.  The correlation function is exponential to a good approximation. Figure~\ref{fig:premeltprop} also shows a representative configuration of the premelting layer and an iso-density surface locating the instantaneous boundary of the coarse-grained order field as calculated from Eq.~\ref{eq:CG}. We define this boundary as being where the field is locally equal to $1/2$ of its average bulk value. This procedure for constructing an instantaneous interface is similar to previous calculations used for calculating liquid-vapor interfacial properties.\cite{willard2010instantaneous} 

The crystal-liquid surface tension determines the mean squared fluctuations in the height of the interface.\cite{nelson1989statistical} The fact that the liquid-crystal surface tension is about half of the liquid-vapor surface tension\cite{eisenberg2005structure} means that while there is a significant free energetic penalty to forming an interface, the size of fluctuations that result at such an interface are large in amplitude. These fluctuations are larger than what are seen at ``soft" liquid-vapor interfaces. The dispersion relation for the size of fluctuations over large lengthscales is the same as that found for capillary waves, $\langle | h(k) |^2 \rangle \sim \kB T/ \gamma k^2$ where $ | h(k) |^2 $ is the amplitude of the Fourier transform of the height-height correlation function, $\kB T$ is Boltzmann's constant times temperature and $ \gamma$ is the surface tension. The coarse-graining procedure used, which evaluates $\bar{q}(\mathbf{r})$ on a grid with resolution of 1 $\mathrm{\AA}$, recovers this scaling relation for wave vectors smaller than 0.6 $\mathrm{\AA}^{-1}$, as shown in Fig.~\ref{fig:premeltprop}c. At low temperatures, away from the large fluctuations at $\Tm$, the surface tension calculated in this way agrees well with previous estimates from free energy calculations. Specifically for the mW model, at $T=255$ K, this procedure yields an estimate of the liquid-crystal surface tension equal to 36 mJ$/$m$^2$, compared to 35 mJ$/$m$^2$ found previously.\cite{limmer2012phase} 

At the temperatures considered here, crystal orientation has a negligible effect on our results.  This fact is important for the field theory model we adopt in the next section.  The insensitivity near the melting temperature to crystal orientation is examined further in subsequent sections.  The insensitivity suggests that the roughening temperatures of ice are lower than the temperatures we consider.\cite{weeks1973structural} 

At the melting temperature, the size of height fluctuations grows due to the interface becoming delocalized. This can be appreciated by the blue lines in Fig.~\ref{equPD:profiles} that plot the distribution of the coarse-grain field, calculated in the same way as in Eq.~\ref{eq:qofz} but with $\bar{q}(\mathbf{r})$ replacing $q(\mathbf{r})$. As expected, these curves are smooth and lay basically on top of the bare order parameter profile. The width of the interface is proportional to the size of these fluctuations, and as can be seen in Fig.~\ref{equPD:profiles} grows upon approaching $\Tm$. To the extent that this coarse-graining procedure provides a map to a simple statistical field theory, the sigmoidal character of these curves away from $\Tm$ is expected. In particular, the inclusion of only a squared gradient interfacial term in such a theory makes it isomorphic with the van der Waals theory of liquid vapor coexistence, a standard result of which is this qualitative density profile.\cite{rowlinson2002molecular} This connection with the coarse-grained field computed here is made quantitative in the next section.

\section{Logarithmic growth approaching $\Tm$}
In order to understand the temperature dependence of the premelting layer, and to test quantitative predictions for experiment, we turn to statistical field theory. We adopt a general phenomenological Hamiltonian for an order-parameter field parameterized with experimental data.\cite{landau1965collected} This approach was used previously to compute the stability of ice-like structures in hydrophilic nanoconfinement\cite{limmer2012phase} and is only briefly reviewed here. Specifically, we expand the order parameter density defined in Eq.~\ref{eq:q6i} up to fourth order, and keep only the lowest order non-trivial gradient term,
\begin{equation}
\label{eq:ham}
\mathcal{H}_\mathrm{s}[q(\mathbf{r})] =k_\mathrm{B} T \int \, \mathrm{d}\mathbf{r} \,  \left[ f(q(\mathbf{r})) + \frac{m}{2} |\nabla q(\mathbf{r}) |^2 \right ]
\end{equation}
where local free energy is,
\begin{equation}
\label{eq:fe}
f(q) = \frac{a_\mathrm{o}}{2}(T-T_\mathrm{s})\,q^2 - w q^3 + u q^4 
\end{equation}
and the expansion parameters $a_\mathrm{o},w,u$ and $m$ are functions of the heat of fusion, surface tension and melting temperature and $\Ts$ is the temperature of liquid stability.\cite{limmer2012phase} While in principle all of these parameters depend on temperature and pressure, such a dependence is neglected here as we consider only conditions of ambient to low pressure, and include only the lowest order temperature dependence.

In order to analyze the premelting transition with this theory, we proceed by making a number of simplifications. First, we recognize on average the system is symmetric in the plane parallel to the interface. Therefore we can trivially integrate out the degrees of freedom in the $x$-$y$ plane. Next, we follow Lipowsky\cite{lipowsky1982critical} and expand the interfacial free energy up to harmonic fluctuations of the order parameter field that acts at the mean location of the end of the slab, $z^*$.  The resultant effective Hamiltonian per unit area is 
\begin{equation}
\label{eq:ham_tot}
\bar{ \mathcal{H}}[q(z)] = \bar{\mathcal{H}}_\mathrm{s}[q(z)] + \frac{k_\mathrm{B} T \,a_\mathrm{s}}{2}\int_{-\infty}^\infty \, \mathrm{d}z \,  q(z)^2\delta(z-z^{*})  \,,
\end{equation}
where $\bar{\mathcal{H}}_\mathrm{s}[q(z)]$ is the Hamiltonian of Eq.~\ref{eq:ham} evaluated for a $z$-dependent $q(\mathbf{r})$, divided by the area of the system in the $x$-$y$ direction. The parameter, $a_\mathrm{s}$, is related to the excess surface tension, and is derivable from lattice models where it relates the average interaction strength at the surface to that in the bulk.\cite{lipowsky1984surface} For reasons that are clarified below, we require only that $a_\mathrm{s}$ is large relative to $a_\mathrm{o}$ and positive.\cite{lipowsky1984surface} The second term on the right hand side of Eq.~\ref{eq:ham_tot} effectively enforces a boundary condition that specifies a disordered region for $z>z^*$.

While it is not analytically tractable to solve for the complete partition function prescribed by this Hamiltonian, we can approximate it by neglecting fluctuations and in doing so compute the mean interface profile. The mean field free energy is given by
\begin{equation}
F_\mathrm{MF}(q) = \bar{\mathcal{H}}(\langle q(z) \rangle) \, ,
\end{equation}
where, 
\begin{equation}
\frac{\delta  \bar{\mathcal{H}}}{\delta \langle q(z) \rangle} = 0 \, ,
\end{equation}
and $\langle q(z) \rangle$ is the order parameter profile that minimizes the effective Hamiltonian. The resultant Euler-Lagrange equation determines the form of the profile,
\begin{equation}
\frac{\dd q}{\dd z} = \left( 2[f(q)-f(q_\mathrm{xtl})]/m \right)^{1/2} \, ,
\end{equation}
with the implicit equation,
\begin{equation}
\left (a_\mathrm{s} q_\mathrm{s}\right )^2 = f(q_\mathrm{s})-f(q_\mathrm{xtl}) \, ,
\end{equation}
for the value of the order parameter at the interface, $q_\mathrm{s}=q(z^*)$. 
\begin{figure}[b]
\begin{center}
\includegraphics{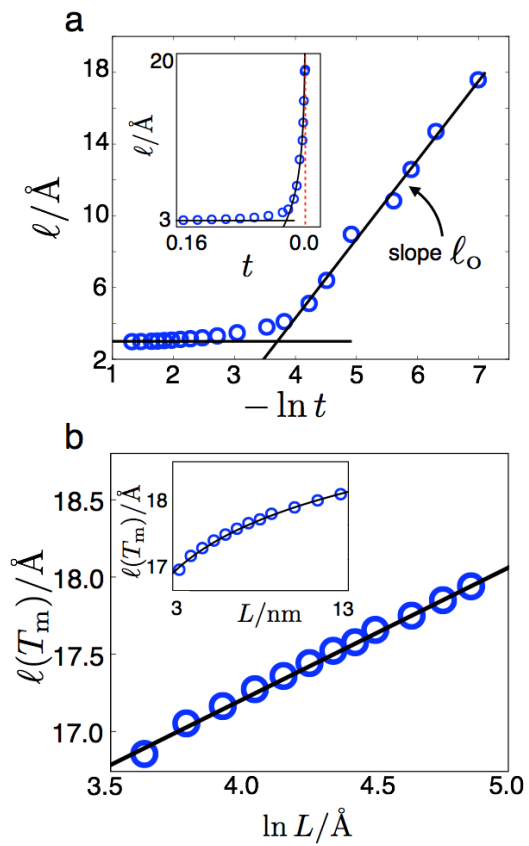}
\caption{Scaling relations for the premelting layer thickness for a) temperature and b) system size. Blue data points locate simulation results for an mW model ice Ic surface exposed to its vapor. Black lines are predicted forms based on the field theory. Insets reproduce the same data on a linear scale. Errors are the size of the symbols. Experimentally, the logarithmic scaling is expected to be valid up to 4-5 nm. See text for details.}
\label{Fi:3}
\end{center} 
\end{figure}

Given the form of the order parameter profile, we can solve for the equation of the thickness of the premelting layer. As given in Ref.~\onlinecite{lipowsky1982critical}, this is a logarithmic function of temperature with a divergence at $\Tm$,
\begin{equation}\label{eq:premelt}
\ell (t)  = - \ell_\mathrm{o}  \ln \left |t \right |+ \mathcal{O}(1) \, \,\,\, \mathrm{for} \,\,\, \, a_\mathrm{s} \geqslant \sqrt{a_\mathrm{o}}
\end{equation}
where $t=(\Tm-T)/\Tm$ and 
\begin{eqnarray}
\ell_\mathrm{o} & = & (1/2) \left[(a_\mathrm{o}/m)(\Tm - \Ts)   \right]^{1/2} \notag \\
 & = & 3 \Delta \gamma / 2 \Delta h \,,
\label{eq:ello}
\end{eqnarray}
where, similar to previous work,\cite{limmer2012phase} the second equality relates parameters to the surface tension difference, $\Delta \gamma$, and enthalpy density difference between liquid and crystal, $\Delta h$. We note that previous work considered only systems with liquid-solid interfaces, thus only $\gamma_{s,l}$ entered into the derived relations rather than $\Delta \gamma$, which includes contributions to the additional solid-vapor and liquid-vapor interfaces present here. The logarithmic divergence of Eq.~\ref{eq:premelt} is typical of surface transitions and is also found in wetting and pinning transitions.\cite{nelson1989statistical} In all cases, interface delocalization is due to close coexistence of multiple bulk phases. The criterion in Eq.~\ref{eq:premelt} explains why premelting layers do not exist on all solids, for when $a_\mathrm{s} < \sqrt{a_\mathrm{o}}$, the surface changes discontinuously at $\Tm$.   

Using the coarse-grained profiles in Fig.~\ref{equPD:profiles}, as well as ones at other temperatures, we can calculate the thickness of the premelting layer as a function of temperature for the mW model. We define this thickness by taking the difference of the Gibbs dividing surface, which marks the decay of the density field, and the analogous location of $\bar{q}(z)=1/2$. Figure~\ref{Fi:3} reports this data. In the semilog plot of Fig.~\ref{Fi:3}a, for reduced temperatures, $t < e^{-2}$, there is a rise in the premelting thickness that crosses over to logarithmic growth for $t < e^{-4}$. Also plotted is the predicted scaling from Eq.~\ref{eq:premelt} using the parameters calculated for the mW model, which determine $\ell_\mathrm{o} =4.6$\AA. This is computed using Eq.~17 with $\gamma_{\ell,s}$,\cite{limmer2012phase} $\gamma_{\ell,v}$,\cite{Molinero:2009p4008} and $\Delta h$\cite{Molinero:2009p4008} computed previously, and $\gamma_{s,v}$ computed from the Kirkwood-Buff virial equation for the 100 surface of mW ice Ic at 200 K,\cite{rowlinson2002molecular} yielding $\Delta \gamma = 82$ mJ/m$^2$.

As shown, the agreement found is within the error of the calculation, suggesting that additional fluctuation effects are not important for the reduced temperatures approached here. The minor role of fluctuations can be anticipated from previous work, which determined $d=3$ as the critical dimension for premelting with a thermal field that is unmodified by fluctuations.\cite{lipowsky1984surface} The low temperature behavior and asymptotic value, $\ell \approx 3\mathrm{\AA}$, likely depend sensitively on the details of algorithm and order parameter and can as seen in Fig.~\ref{equPD:profiles} is not expected to be well predicted from this type of field theory.

Figure~\ref{Fi:3}b also shows the scaling of the premelting thickness at the melting temperature, $\ell(\Tm)$, with the linear dimension of the system, $L$. Finite-size scaling arguments predict that the premelting length at the melting temperature under our inhomogeneous boundary conditions should scale as $\ell(\Tm) \sim \ln \,L$.\cite{lipowsky1984interface}  This scaling is indeed found for the limited system sizes that we can simulate. Figure~\ref{Fi:3}b plots data taken from different sized systems, characterized by increasing $L$ with $W$ fixed. The black line is a fit with a correlation constant that is within the uncertainty of the data. This logarithmic scaling can be rationalized by considering that the coexistence conditions for a finite system change in proportion to the ratio of surface to volume, which in this case is proportional to $1/L$. The argument in the logarithmic function in Eq.~\ref{eq:premelt} for the thickness depends on these coexistence conditions, therefore it is not surprising that this scaling, $-\ln 1/L$, exists. 

It is important to note that simulations with the mW model and the theory described above omit a proper accounting for long-ranged interactions. For thick premelting layers, such long-ranged forces have been suggested to dominate the observed scaling behavior.\cite{elbaum1991application} The effect of long-ranged interactions can be estimated by computing the effective Hamaker constant for a system of ice, liquid water and vapor.\cite{israelachvili2011intermolecular} By including only contributions of the static dielectric constants into DLP theory,\cite{dzyaloshinskii1961general} we obtain a value of -0.026 $\kB T$ for this constant. It is negative because of the dielectric constant of liquid water lies between those of ice and vapor.  It is small in magnitude because there is only a 4\% difference between dielectric constant of water and ice at the melting temperature.\cite{eisenberg2005structure} The negative sign implies an attraction between slabs of ice and vapor mediated through the liquid. The attraction implies a finite pre melting layer thickness, rather than a divergence, as $t \rightarrow 0$. This incomplete wetting behavior\cite{tartaglino2005melting} has been previously inferred from SFG experiments.\cite{wei2001surface} We can account for this effect in our mean field theory by adding an asymptotically correct term for the dispersion interaction that scales as $\ell^{-2}$.  This term allows us to predict a plateau from logarithmic growth at $t \approx 10^{-5}$ and an ultimate pre melting thickness of $\ell(t = 0) \approx 6$nm. As such, we predict that results for the mW model and the theory in Eq. 11 are accurate relative to experiment up to thicknesses of around 4-5 nm.

This dominance of the logarithmic growth away from $\Tm$ allows us to use Eq.~16 to make a reasonable estimate for the thickness of the premelting layer expected for experiment. In particular, we can use experimental parameters for the heat of fusion and surface tension,\cite{eisenberg2005structure} to determine $\ell_\mathrm{o}=5.2 \mathrm{\AA}$. Assuming the same $\mathcal{O}(1)$ constant found for the mW model, Eq.~\ref{eq:premelt} determines $\ell$ at 272~K for experiment to be 3~nm. This value is in good agreement with near-edge X-ray absorption fine-structure spectra that reports $\ell(272 \,\mathrm{K})=3$~nm\cite{bluhm2002premelting} and reasonable agreement with ellipsometry measurements that reports $\ell(272\,  \mathrm{K})=5$~nm.\cite{beaglehole1980transition} 

\section{Universal premelting on ice}
The field theory and its consequences discussed in the previous sections are general. The form for the energy that controls the collective excitations, like interface formation and deformation, is dictated entirely by the symmetry of the Hamiltonian. The experimental observables that determine the actual values of these excitations and their likelihood are determined from macroscopic observables. Therefore, to the extent that the properties entering into the parameterization of Eqs.~\ref{eq:ham} and \ref{eq:fe}, do not change, then the behavior of the interface will not be sensitive to details such as the crystal orientation or bulk lattice. Indeed, experimental estimates of the heat of fusion and surface tension for different crystalline lattices or planes do not vary by more than 10$\%$.\cite{kuroda1982growth}

\begin{figure}
\begin{center}
\includegraphics{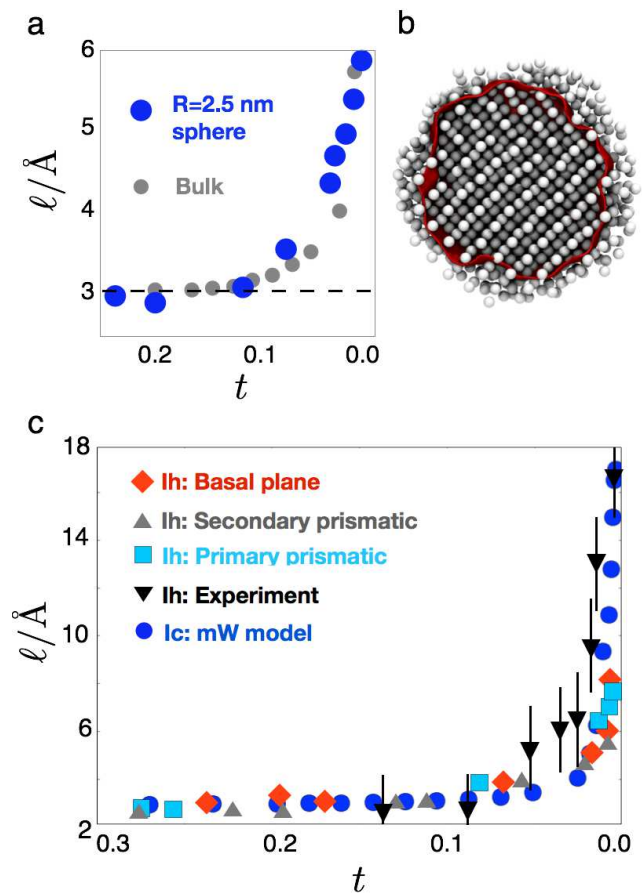}
\caption{Universality of premelting on surfaces of ices. a) Premelting thickness on the surface of a nanocrystalline sphere compared to the bulk premelting thickness. Note the $x$-axis accounts for melting point depression for the finite sized sphere. b) A cut through a representative configuration of a spherical crystallite and its instantaneous order parameter surface. c) Premelting thickness for different orientations of ice Ih taken from Ref.~\onlinecite{conde2008thickness}, compared to experimental results from Ref.~\onlinecite{bluhm2002premelting} and our results from the mW model. As in a) the $x$-axis accounts for the different melting temperatures of the different models and lattices.}
\label{Fi:5}
\end{center} 
\end{figure}

Figure~\ref{Fi:5} confirms this expectation. Specifically we plot the premelting layer thickness for ice Ih along three different crystal facets and a spherical crystal with radius 3 nm. Data from ice Ih has been taken from Ref.~\onlinecite{conde2008thickness}, which used the TIP4P/2005 model with MD simulations performed in an analogous way to our results for the mW model. These simulations did not use a coarse-grained order parameter profile to determine the premelting length, however in plotting it here we assume the scaling holds for their calculation up to an additive constant, which has been checked explicitly with our own calculations for the basal plane. This constant is $\mathcal{O}(1)$ and depends slightly on the lattice orientation. Data for premelting on the surface of an ice sphere was generated with the mW model. To create this surface, a sphere of radius 2.5 nm was cut out of a crystalline ice Ic lattice. Prior to simulation, the surface was relaxed by removing molecules coordinated to fewer than 3 neighbors as determined by a radial cutoff of 3.3 \AA. 

In each case presented in Fig.~\ref{Fi:5}, a premelting layer exists and grows as the melting temperature is approached. Qualitative details such as the basic form of the temperature dependence and magnitude are conserved, while quantitative details such as the effective, $\ell_\mathrm{o} $, governing the temperature dependence are sensitive to orientation and bulk lattice. In principle, these differences can be accounted for by computing $\Delta \gamma$, $\Delta h$, and $T_\mathrm{m}$ for each of these crystals and orientations. For the curved spherical surface, the finite system dictates a small maximal premelting thickness, $\sim \log R$. The finite size of this system also dictates that curvature corrections and fluctuation effects broaden the transition region, which can be similarly appreciated in Fig.~\ref{Fi:5}. Such finite size effects also explain observations of a near constant thickness disordered layer of water in hydrophilic confinement.\cite{alba2006effects} Indeed the success of simple theories for freezing in confined systems relies on the existence of this premelting layer and its negligible temperature dependence for systems whose size is large enough to produce measurable shifts in the melting temperature.\cite{limmer2012phase} These simulation results also agree particularly well with the experimental results of Bluhm et al.\cite{bluhm2002premelting} that are plotted in Fig.~\ref{Fi:5}c.

In some instances, for finite systems, it is possible to arrange molecules at the surface in such a way that hydrogen bonds are not broken. These are systems, such as water confined to hydrophobic nanotubes, $R<20$\AA,\cite{koga2001formation} in which the interface is highly curved. In these instances, a premelting layer is not expected nor observed, as there is no longer an energetic penalty associated with breaking strong cohesive interactions.

\section{Implications of a soft interface}
The physical properties of the surface of ice determine a number of important dynamical processes. In the atmosphere, ice particles are the substrates on which much of atmospheric chemistry occurs. The premelting layer is thought to be especially important in acting as a solvent for acid base reactions.\cite{dash2006physics} The results presented here will help rationalize trends in such reactions. 

Another simple kinetic process that occurs on the surface of ice is evaporation. Experimentally, it is known that the evaporation of a water molecule from the surface of ice is effectively barrierless between the temperature range of 245 K to 273 K.\cite{sadtchenko2004vaporization} This is inferred by the activation energy being equal to the thermodynamic heat of sublimation, as determined by measuring the rate of evaporation as a function of temperature with microcalorimetry and mass spectrometry.  Recent work on the evaporation from the surface of liquid water that has concluded that this process is similarly barrierless.\cite{varilly2012water} It would seem that the existence of a disordered interface at the boundary between the condensed phase and the vapor supplies sufficiently facile reorganization around an evaporating molecule that evaporation from either a liquid or a quasi-liquid layer is simply a ballistic process. Only for low temperature, where the premelting layer vanishes, $T\lesssim245$~K, is evaporation expected to have to overcome a barrier.

In this work, we have used a simple field theory supplemented by atomistic simulations, coarse-graining and experimental data to analyze the structure of ice interfaces. We find generically that near the bulk melting temperature, there exists a premelting layer whose thickness changes continuously away from coexistence. While the simulations and theory assume that short-ranged, molecular interactions dominate, which leads to the characteristic logarithmic scaling of the premelting layer thickness with temperature, the effects of long-ranged interactions have been estimated and found to be relevant only at small reduced temperatures, $t<10^{-5}$ ultimately producing a finite premelting thickness of 6nm. We have made an estimate of the expected premelting thickness at 272 K of 2 nm, which is in agreement some experimental results\cite{bluhm2002premelting,beaglehole1980transition} but not others.\cite{golecki1978intrinsic,dosch1995glancing}

 \begin{acknowledgments} 
The authors thank Gabor Somorjai for pointing us to the phenomenon of premelting and Erio Tosatti for insightful comments. In its early stages this research was supported by the Helios Solar Energy Research Center of the U.S. Department of Energy under Contract No. DE-AC02-05CH11231. In its final stages, it was supported by the Director, Office of Science, Office of Basic Energy Sciences, Materials Sciences and Engineering Division and Chemical Sciences, Geosciences, and Biosciences Division under the same DOE contract number. DTL is supported as a fellow of the Princeton Center for Theoretical Science. 
\end{acknowledgments}

%


\end{document}